\input harvmac
\input epsf
\input rotate
\input xyv2


\def\dnl{D_{n,l}}
\def\ep{\epsilon}

\def\rp{\rho_\phi}
\def\rg{\rho_\gamma}
\def\g{\Gamma}
\def\w{\omega}


\Title{USTC-ICTS-07-16}{Reheating and Cosmic String Production}


\centerline{Chao-Jun Feng$^{1,2}$, Xian Gao$^{1,2}$, Miao
Li$^{1,2}$, Wei Song$^{1,2}$, Yushu
Song$^{1,2}$\footnote{$^\dagger$} {fengcj@itp.ac.cn,
gaoxian@itp.ac.cn, mli@itp.ac.cn, wsong@itp.ac.cn,
yssong@itp.ac.cn}}
\bigskip

\centerline{\it $^1$ Interdisciplinary Center of Theoretical
Studies}
\centerline{\it USTC, Hefei, Anhui 230026, China}

\bigskip

\centerline{\it and}

\bigskip

\centerline{\it $^2$ Institute of Theoretical Physics}
\centerline{\it Academia Sinica, Beijing 100080, China}


\nref\rrvi{J. Schwarz, hep-th/0702219.}
\nref\polre{
  J.~Polchinski,  AIP Conf.\ Proc.\  {\bf 743}, 331 (2005)
  [Int.\ J.\ Mod.\ Phys.\  A {\bf 20}, 3413 (2005)], hep-th/0410082.}
\nref\oldcs{A. Vilenkin and E. Shellard, {\it Cosmic Strings and
Other Topological Defects,} Cambridge University Press (Cambridge,
2000); M.Hindmarsh and T. Kibble, Rep. Prog. Phys. {\bf58}, 477
(1995).}
\nref\witten{E. Witten, Phys. Lett. {\bf B153}, 243 (1985).}
\nref\jst{N. Jones, H. Stoica, and S. Tye, JHEP {\bf 07}, 051
(2002), hep-th/0203163.}
\nref\st{S. Sarangi and S. Tye, Phys. Lett. {\bf B536}, 185 (2002),
hep-th/0204074.}
%
%
\nref\gubser{S. Gubser, Phys.Rev. {\bf D69} (2004) 123507,
hep-th/0305099; hep-th/0312321.}
\nref\lss{M. Li, W. Song and Y. Song, hep-th/0701258, JHEP 04 (2007)
042.}
\nref\gsw{M. Green, J. Schwarz and E. Witten, {\it Superstring
Theory,} Vols. I and II, Cambridge Univ. Press (1987).}
\nref\sdcm{E. Brezin and C. Itzykson, Phys. Rev. {\bf D2} (1970)
1191; D. Chung, hep-ph/9809489.}
%

\bigskip

\medskip
\noindent
We compute the string production rate at the end of inflation, using
the string spectrum obtained in \lss\ in a near-de Sitter space. Our
result shows that highly excited strings are hardly produced, thus
the simple slow-roll inflation alone does not offer a cosmic string
production mechanism.


\Date{June 2007}


\newsec{Introduction}

String theory has been successful in resolving some longstanding
problems, such as the existence of a consistent theory of quantum
gravity. However, many problems remain unsolved \rrvi. One of the
most important problems is that by far string theory has not made
any concrete predictions verifiable by experiments, thus we do not
know whether string theory is a realistic physical theory or not.
Cosmology may be an important arena to test string theory. For
example, any evidence of the existence of topological defects such
as cosmic superstrings, can be an important support for string
theory. Cosmic strings can have two different origins, the field
theoretic one and fundamental string theory. The discovery of
fundamental cosmic strings would be a spectacular way to verify
string theory \polre. Cosmic strings from string theory are
characterized by some properties not shared by GUT cosmic strings
\oldcs.

In 1980's, it was generally believed that the perturbative
fundamental strings can not become cosmic strings, due to an
argument of Witten \witten . Moreover, the tension of a fundamental
string is close to the Planck scale, while cosmic strings with such
a tension are ruled out by experiments. Recently, research of
compactifications in string theory shows that the string tension
measured in the four-dimensional Einstein frame can be much smaller,
and in some situations the instability problem is evaded. Thus
cosmic strings as fundamental strings may indeed exist and can be
observed in the future experiments.

The current research interest of the creation of strings at the end
of inflation is focused on the investigation of the final results of
collision of branes. Cosmic strings are inevitably produced in this
process as topological defects. Research along this direction is
spearheaded by Polchinski and Tye, and their collaborators \jst \st
\polre.

In this paper, we will study creation of strings in a more
traditional fashion, namely, through gravitational pair production
in a time-dependent background. Related work has been done by Gubser
\gubser\ with an effective field theory viewpoint, and a steepest
descent contour method has been developed to estimate the production
rate of the strings.

As discussed in \gubser, in a regime of parameter space where a
spacetime description gives a good approximation of string dynamics,
the on-shell constraint for a given string state boils down to a
differential equation describing an oscillator with a time-dependent
frequency. When the quantization of strings is carried out in a
particular background, the ``frequency" $\omega(t)$ is determined.
The quantization of strings in a de Sitter background was recently
done by Li et al. \lss, and the spectrum of such ``small strings" is
obtained. We will use the method developed in \gubser\ to estimate
the total rate of string creation, using the spectrum i.e. the
equation of motion of string state obtained in \lss, which is
different from that of \gubser. The original equation of motion
derived in \lss\ is in de Sitter space where the Hubble parameter is
a constant, but as we will show that in fact it is also valid at the
end of inflation and during reheating, where the Hubble parameter is
a function of time instead of a constant (see Appendix A). We will
show this in Appendix B.

The main result of our investigation can be summarized as follows.
Strings are generally produced gravitationally at the end of
inflation and during reheating, and the energy density of strings
produced is highly suppressed by an exponential factor multiplied by
an power factor. Since the Hubble parameter is much smaller than 1
in the unit of $\alpha'=1$, this energy density is very small. Our
estimate is approximate quantitatively, due to the fact that there
are some approximations used in deriving the string spectrum derived
in \lss\ and in our analytic method in estimating the string
production rate. However, this semi-quantitative result strongly
suggests to us the picture that highly excited strings are hardly
produced during reheating and the production rate is very small.

The organization of this paper is as follows. In section 2 the
density of string states is calculated, based on the spectrum
obtained in \lss. In section 3 the creation rate and the energy
density of strings are estimated. The final section is devoted to
discussions. In Appendix A we solve the Friedmann equation directly
to get the Hubble parameter as a function of time at the end of
inflation and during reheating. In Appendix B We show that the
spectrum and equation of motion of strings obtained in \lss\ are
valid in a general flat FRW background, not only in pure de Sitter
space.

\newsec{Degeneracy of String States}

The spectrum of strings in a near de Sitter background is different
from the one in flat spacetime, and depends on two integers (we
shall consider the bosonic sector only in this paper), as will be
shown in the next section. These integers are eigenvalues of two
operators: the number operator $N$ and the other operator $L$,
defined respectively as follows
\eqn\NandL{\eqalign{&N=\sum_{i=1}^{d}\sum_{m=1}^{\infty}m\left(N^i_m+\tilde
N^i_m\right)\cr
&L=\sum_{i=1}^{d}\sum_{m=1}^{\infty}\left(N^i_m+\tilde N^i_m+2N^i_m
\tilde N^i_m\right),\cr}}
where $d=D-1$ and $D$ is the number of dimensions of spacetime,
$N^i_m$ and $\tilde N^i_m$ are the occupation numbers of the
left-mover and the right-mover respectively, $i$ is the space index
and $m$ is the oscillator index. We denote $n$ as the eigenvalue of
$N$ and $l$ as the eigenvalue of $L$. The degeneracy of states at
level $n$ and fixed $l$ is denoted by $\dnl$, which is encoded in a
generating function as the coefficient of $z^nw^l$
\eqn\pa{Z(z,w)\equiv trz^Nw^L,}
or
\eqn\gen{Z(z,w)=\sum_{n,l=0}^{\infty}\dnl z^nw^l.}
Using \NandL\ and \pa\ , the generating function can be evaluated by
an elementary method of quantum statistical mechanics as follows
\eqn\generatingorign{Z(z,w)=\prod_{i=1}^{d}\prod_{m=1}^{\infty}\sum_{N^i_m,\tilde{N}^i_m=0}^{\infty}z^{m(N^i_m+\tilde
N^i_m)}w^{N^i_m+\tilde N^i_m+2N^i_m \tilde N^i_m}.}
Summing over \foot{Of course one can sum over $N^i_m$ first.}
$\tilde N^i_m$
\eqn\generating{\eqalign{Z(z,w)&=\prod_{i=1}^{d}\prod_{m=1}^{\infty}\sum_{N^i_m=0}^{\infty}{z^{mN^i_m}w^{N^i_m}\over1-z^mw^{2N^i_m+1}}\cr
&=\prod_{i=1}^{d}\prod_{m=1}^{\infty}\sum_{N^i_m=0}^{\infty}{z^{mN^i_m}\over1-z^m}f(w,
N^i_m),\cr}}
where
\eqn\funf{f(w, N^i_m)={(1-z^m)w^{N^i_m}\over1-z^mw^{2N^i_m+1}}.}
If $w=1$, \generating\ is just the usual generating function of the
degeneracy of bosonic string states \gsw. The effect of $w$ is to
deform this generating function. This deformation is small when it
comes to evaluate the coefficient $\dnl$ by the steepest descent
method. We will expand the generating function i.e. \funf\ near
$w=1$ to the second order, with higher order terms truncated. This
assumption is reasonable as we will show in the saddle point
calculation.
\eqn\expfunf{f(w,
N^i_m)=1+(w-1)\left[{z^m\over1-z^m}+N^i_m\left({1+z^m\over1-z^m}\right)\right]+O((w-1)^2).}
Thus the summation in \generating\ reads:
\eqn\partw{\sum_{N^i_m=0}^{\infty}{z^{mN^i_m}\over1-z^m}f(w,
N^i_m)\approx{1\over(1-z^m)^2}\left[1+(w-1){2z^m\over(1-z^m)^2}\right],}
then the generating function is
\eqn\partf{Z(z,w)\approx\left\{\prod_{m=1}^{\infty}{1\over(1-z^m)^2}\left[1+(w-1){2z^m\over(1-z^m)^2}\right]\right\}^d.}

Let $z\equiv\exp(-{1\over T})$ and $x\equiv{{ m\over T}}$. Thus $x$
is continuous when $T$ is large enough. In our following
calculation, around the saddle points, $z$ is close to 1 thus $T$ is
indeed large. The infinite product in
\partf\ is approximated by an exponential of an integral. Taking logarithm of both sides of
\partf\ we obtain
\eqn\logparti{\eqalign{ {\ln Z(z,w)\over d}
&\approx-{2T}\int_{{1\over
T}}^{\infty}dx\ln(1-e^{-x})+T\int_{{1\over
T}}^{\infty}dx\ln\left(1+(w-1){2e^{-x}\over(1-e^{-x})^2}\right)\cr
&\approx{\pi^2T\over3}+{2(w-1)T}\int_{{1\over
T}}^{\infty}{dxe^{-x}\over(1-e^{-x})^2}\cr &=-{\pi^2\over3\ln
z}-{2(w-1)z\over(1-z)\ln z} ,\cr}}
in the second line we have truncated higher orders of $w-1$ in the
logarithm function.

By definition, the degeneracy of string states is
\eqn\density{\dnl=\oint {dw \over2\pi i} \oint {dz \over2\pi
i}{Z(z,w)\, \over z^{n+1}\, w^{l+1}.}}
Here $Z(z,w)$ vanishes rapidly as $z\rightarrow1$ when $w<1$, and
$z^{n+1}$ is very small for $z<1$ when $n$ is very large.
Consequently, for large $n$, there is a sharply defined saddle point
for $z$ near $1$. Indeed, the factor \foot{Here $d=2$ is the
dimension of physical states, i.e. the transverse oscillators are
$N_m^i, i=1,2$.}
\eqn\factor{\exp[-{2\pi^2\over3\ln z}-{4(w-1)z\over(1-z)\ln
z}-(n+1)\ln z-(l+1)\ln w] }
is stationary for
\eqn\saddle{\ln z=-\sqrt{{1\over
n+1}\left({2\pi^2\over3}-{8(w-1)\over\ln z}\right)} \sim
-\sqrt{{1\over n+1}{2\pi^2\over3}.}}
Therefore one finds that as $n\rightarrow\infty$
\eqn\densityn{\dnl\sim n^{-5/4}\,
\exp\left({2\pi\over3}\sqrt{6n}\right)\,
\delta\left(l-{6n\over\pi^2}\right).}
where the $\delta$-function comes from the integration over $w$, and
the exponential factor is the ordinary degeneracy of bosonic string
states in four-dimensional spacetime.

\newsec{Energy Density of Cosmic Strings}

In this section we will use the steepest descent contour method
developed in \gubser\sdcm\ to estimate the energy density of strings
produced during reheating. In \lss, the quantization of bosonic
strings has been done in a de Sitter background. The on-shell
constraint for quantum states of a string leads to an equation of
the form:
\eqn\mlss{\eqalign{\Big\{
&\partial_t^2+3H\partial_t+k^2e^{-2Ht}+4N+2E_0\cr
&-\sum_{m,i}H^2\,\left(1+2N^i_{m}\,\tilde{N}^i_{m}+N^i_{m}+\tilde{N}^i_{m}\right)\Big\}
\phi(N^i_m,\tilde{N}^i_m,\omega,k^i)=0,}}
where we take $\alpha'=1$. $H$ is the Hubble parameter, $N^i_m$,
$\tilde{N}^i_m$ are occupation number operators and $N \equiv
\sum_{i,m} m (N^i_m + {\tilde{N}}^i_m)$, $k^i$ is the momentum
vector in the four-dimensional spacetime and $E_0$ is the
center-of-mass energy. A general physical state $|\phi\rangle$
corresponding to the string modes is related to
$\phi(N^i_m,\tilde{N}^i_m,\omega,k^i)$ as follows,
    \eqn\state{|\phi\rangle = \sum_{N^i_m,\tilde{N}^i_m} |N^i_m,\tilde{N}^i_m,\omega,k^i\rangle
    \phi(N^i_m,\tilde{N}^i_m,\omega,k^i),}
where the definition of $|N^i_m,\tilde{N}^i_m,\omega,k^i\rangle$ can
be found in \lss. Eq.\mlss\ is regarded as the equation of motion
for the field of the corresponding string state. As we mentioned in
the introduction, this equation is different from that used in
\gubser, this difference makes our new result different from others.

There is no string production in pure de Sitter space with a
constant Hubble parameter, even with the modified string spectrum as
in \mlss. The only chance for string production to occur is the
short period of reheating during which $H$ becomes time-dependent.
Thus we have to make a step forward, i.e. to generalize the equation
of motion \mlss\ to the case when $H$ varies with time. This is
developed in Appendix B. From now on $(k^i)^2e^{-2Ht}$ in \mlss\ is
replaced by $(k^i)^2/a(t)^2$, where $a(t)$ is the cosmological scale
factor.

It is convenient to introduce $\phi(t)$ via $\Phi(t) \equiv a
\phi(t)$, thus the equation of motion for $\phi(t)$ is:
\eqn\multbya{\ddot \phi+H\dot\phi+\left[\left({k^i\over
a(t)}\right)^2+4N+2E_0-C^2H^2\right]\phi=0}
where
\eqn\csqure{C^2\equiv\sum_{m,i}\left(3-\ep+N^i_m+\tilde N^i_m+2N^i_m
\tilde N^i_m\right)} and
\eqn\epsl{\ep\equiv-{\dot{H}\over H^2}}
defined as the so-called slow-roll parameter, which is roughly equal
to $1$ at the end of inflation, and the dot denotes the derivative
with respect to $t$. The slow-roll parameter is not larger than $2$
in the case under study, see Appendix A.

In conformal time $\eta$ defined by $ad\eta=dt$, we can eliminate
the first order derivative term, and we will use prime to denote the
derivative with respect to the conformal time. Thus \multbya\
becomes:

\eqn\eomaftercon{\phi(\eta)''+W(\eta)^2\phi(\eta)=0,}
where
\eqn\kandm{W(\eta)^2 \equiv
k^2+(4N+2E_0)a(\eta)^2-C^2a(\eta)^2H(\eta)^2.}

Having obtained the equation of motion \eomaftercon\ , now we use
the steepest descent method to extract the approximate string pair
production rate from \eomaftercon\ . The steepest descent method was
developed by various authors, especially, it was used to estimate
the string production rate by Gubser \gubser\ . The key assumption
is that the occupation number $|\beta|^2$ for a given mode is always
much less than 1, where $\beta$ is the Bogliubov coefficient.
Setting
\eqn\AlphaBeta{\phi(\eta) = {\alpha(\eta) \over
\sqrt{2W(\eta)}}e^{-i \int^\eta du \, W(u)}+{\beta(\eta) \over
\sqrt{2W(\eta)}}e^{i \int^\eta du \, W(u)} \,,}
with the requirement $|\alpha(\eta)|^2-|\beta(\eta)|^2=1$, we recast
the equation \eomaftercon\ into
\eqn\abform{\alpha'(\eta) = { W' \over 2W}e^{2i\int^\eta du \,W(u)}
\beta(\eta) \qquad \beta'(\eta) = { W' \over 2W}e^{-2i\int^\eta du
\, W(u)} \alpha(\eta) \,.}
Using the assumption $\beta(\eta) \ll 1$ and $\alpha(\eta) \approx
1$, we obtain an approximate formula for $\beta$
\eqn\betform{\beta\approx\int_{-\infty}^\infty d\eta \, { W' \over
2W}\exp\left( -2i \int^\eta du \, W(u) \right) \,.}

The integral in the exponential of \betform\ can be calculated as
follows
\eqn\expno{\int^\eta_{\eta_i}du
W(u)=\left[\int^{\eta^\star}_{\eta_i}+\int^\eta_{\eta^\star}\right]du
W(u)}
where $\eta_i$ is some initial time and $\eta^\star$ is defined to
make $W(\eta^\star)=0$. Here the second term on the right hand side
of \expno\ can be calculated as follows:
\eqn\expnoI{\eqalign{\int^\eta_{\eta^\star}du
W(u)&=\int^\eta_{\eta^\star}du\sqrt{k^2+(4N+2E_0)a(\eta)^2-C^2a(\eta)^2H(\eta)^2}\cr
&\approx\int^\eta_{\eta^\star}du\sqrt{[(4N+2E_0)2aa'-C^22aa'H^2-C^2a^22HH'](u-\eta^\star)}\cr
&={2\over3}\delta^{3/2}\sqrt{(4N+2E_0)2aa'-C^22aa'H^2-C^2a^22HH'}}}
where we have expanded terms in the square root around $\eta^\star$,
and defined $\delta\equiv\eta-\eta^\star$. Thus
\eqn\betanow{\beta\approx I_0\exp(-2i\int^{\eta^\star}_{\eta_i}du
W(u))}
where \eqn\betaI{I_0 \equiv \int_{-\infty}^\infty d\eta \, { W'
\over 2W}\exp(-{4i\over3}\delta^{3/2}\sqrt{S(\eta^\star)})}
and
\eqn\Sform{S(\eta^\star)=(4N+2E_0)2aa'-C^22aa'H^2-C^2a^22HH'.}
Expanding $W(\eta)$ around $\eta^\star$, we get
\eqn\betaII{I_0={1\over4}\int_{-\infty}^\infty {d\delta\over\delta}
\, \exp\left(-{4i\over3}\delta^{3/2}\sqrt{S(\eta^\star)}\right).}
From \sdcm\ we know that integrals such as \betaII\ can be
calculated by the contour integral method, and the result for
\betaII\ is simply $I_0=i\pi/3$.

Now \betanow\ can be written as:
\eqn\betawithi{\beta\approx{i\pi\over3}\exp\left(-2i\int^r_{\eta_i}du
W(u)\right)\exp\left(-2i\int^{\eta^\star}_rdu W(u)\right)}
where $r$ is the real part of $\eta^\star \equiv r-iu$, and $r, u$
are real with $u>0$ . Since what we need is the modulus of $\beta$,
the main contribution comes from the second exponential function in
\betawithi\ whose argument is the following integral and can be
expanded as:
\eqn\betaint{\int^{\eta^\star}_rdu
W(u)=W(r)(-iu)+W'(r){(-iu)^2\over2}+W''(r){(-iu)^3\over6}+\cdots .}
As long as $|{W''\over W}|\ll|{6\over u^2}|$, we can truncate this
expansion to the first term (even terms do not contribute to the
modulus of $\beta$ since they are real), and indeed in the following
calculation one will see that $|{W''\over W}|\ll|{6\over u^2}|$ is
satisfied in our case. Thus we get
\eqn\betamodulusfinall{|\beta|^2\approx\left({\pi\over3}\right)^2\exp\left(-4uW(r)\right).}
%
Here the imaginary part of $\eta^\star$ can be derived by using
\kandm\ and expanding $W(\eta^\star)$ around $r$ as follows\foot{We
truncate the expansion up to the fourth term because one can check
that the next order terms are much smaller than these terms. And one
can see that the condition $|{W''\over W}|\ll|{6\over u^2}|$ is also
satisfied.}
\eqn\expandw{\eqalign{0 \equiv &
W^2(\eta^\star)=W^2(r)+2W(r)W'(r)(-iu)+\left[W(r)W''(r)+W'^2(r)\right](-iu)^2\cr
&+\left[W'''(r)W(r)+W'(r)W''(r)+2W'(r)W''(r)\right]{(-iu)^3\over3}}}

We solve these equations as follows
\eqn\realandimage{\eqalign{&W^2(r)-[W(r) W''(r)+ W'^2(r)]u^2=0\cr &
6W(r) W'(r)-[W'''(r)W(r)+3W''(r)W'(r)]u^2=0}.}
Then
\eqn\betawithw{|\beta|^2\approx\left({\pi\over3}\right)^2\exp\left({-4W(r)\over\sqrt{W''(r)/
W(r)+\left(W'(r)/ W(r)\right)^2}}\right).}
%
%
In our case $W(\eta)$ is expressed in \kandm, thus we get
\eqn\primewover{{W'\over W}={a'\over a}-\left(k^2({a'\over
a})+C^2[{a''\over a}-2({a'\over a})^2]\right)/W^2\approx{a'\over
a},}
and
\eqn\dobpri{{W''\over W}=\left({W'\over W}\right)'+\left({W'\over
W}\right)^2\approx{a''\over a}.}
Thus we write approximately:
\eqn\betaourcase{|\beta_k(n,l)|^2 \approx \exp\left\{ {-4\left(
{k^2/ a_r^2} +4n - a_r^2 H_r^2 l\right)\over \sqrt{4n\left(
H_r^2+R_r / 6\right)} } \right\},}
%
where we wrote $\beta$ as function of comoving momentum $k$ and
excitation modes $n$, $l$ explicitly, where $n$ and $l$ are
eigenvalues of operators in \NandL\ respectively. We have dropped
the factor of $(\pi/3)^2$, $H_r$ and $R_r$ correspond to the Hubble
expansion rate and Ricci scalar for the metric $ds^2 =
a(\eta)^2(d\eta^2-d{x^i}^2)$ respectively. We emphasize that all
time-dependent quantities in \betaourcase\ are evaluated at $\eta =
r$, where $r$ is the real part of $\eta^{\star}$ given by
$W(\eta^\star)=0$. From \betaourcase\ we can see that indeed the
production of highly-excited strings i.e. strings with large $n$ and
$l$ are exponentially suppressed.
%

\bigskip
The total energy density of strings produced may be written as
\eqn\energydensity{\rho(\eta) = {1\over 2\pi^2 {a(\eta)}^3}\int
{dk}\, k^2 \,\sum_{n,l} \,\dnl\, |\beta_k(n,l)|^2 \, M_{n,l}(\eta),}
where $k$ is the comoving momentum, $M_{n,l}(\eta)$ is the energy of
a single string with excitation modes $(n,l)$, given by
\eqn\energy{\eqalign{M_{n,l}(\eta)^2 &= 4N+2E_0 -
\sum_{m,i}H^2(1+2N^i_m\tilde{N}^i_m+N^i_m+\tilde{N}^i_m)\cr
&=4n+2E_0-H^2(1+l)\cr & \approx 4n-H^2 l}.}
Now insert \betaourcase\ and \energy\ into \energydensity\ , we
obtain approximate formulas:
\eqn\energyden{\eqalign{\rho(\eta) &\approx {1\over 2\pi a^3} \int
dk k^2 \sum_{n,l} n^{-{5\over 4}}\exp\left({2\pi
\over3}\sqrt{6n}\right)\delta(l-{6n\over\pi^2}) \times \cr
&\qquad\qquad \exp\left( -4{k^2/a_r^2+4n-a_r^2H_r^2l \over
\sqrt{4n(H^2_r+R_r/6)}} \right) \sqrt{4n-H^2 l}\cr & \sim  {1\over
a^3} \int dk k^2 \sum_{n} n^{-{5\over 4}}\exp\left({2\pi
\over3}\sqrt{6n}\right) \times\cr &\qquad\qquad \exp\left(
-4{k^2/a_r^2+4n-a_r^2H_r^2{6n\over \pi^2} \over
\sqrt{4n(H^2_r+R_r/6)}} \right) \sqrt{4n-H^2 {6n\over \pi^2}} }.}
For highly excited string states, neglecting $k$ is a good
approximation, though not a uniform one if $a(\eta)$ becomes
arbitrary small in the past. We get:
\eqn\rhofin{\eqalign{\rho(\eta) &\sim {1\over a^3} \int dk k^2
\int_1^{n_{max}} dn n^{-{3\over 4}} \exp \left\{ -\left( {8-12 a_r^2
H_r^2/\pi^2 \over \sqrt{H^2_r +R_r/6} } - {2\sqrt{6}\pi \over 3}
\right) \sqrt{n} \right\} }.}
where we have dropped the constant factor. We define $A=\left({8-12
a_r^2 H_r^2/\pi^2 \over \sqrt{H^2_r +R_r/6} } - {2\sqrt{6}\pi \over
3}\right)$ for short, the above integral can be approximated as:
\eqn\rhoappro{\eqalign{ \rho(\eta) &\sim {1\over a^3 A} \int dk k^2
\left( Erf \left( \sqrt{A} n_{max}^{1\over4} \right) - Erf \left(
\sqrt{A} \right) \right)},}
where $Erf(x)$ is the error function. For very large $x$,
approximately we have
\eqn\errorfunction{Erf(x) \approx 1-
{{e^{-x^2}}\over{x\sqrt{\pi}}}\left [1+\sum_{n=1}^\infty (-1)^n
{{1\cdot3\cdot5\cdots(2n-1)}\over{(2x^2)^n}}\right ] \sim 1-
{{e^{-x^2}}\over{x\sqrt{\pi}}}.}
Thus we have
\eqn\final{\rho(\eta) \sim {1\over a(\eta)^3 A^{3/2}} \left( e^{-A}
- {e^{-A\sqrt{n_{max}}} \over n_{max}^{1/ 4}} \right),}
where $A$ is given by $A=\left({8-12 a_r^2 H_r^2/\pi^2 \over
\sqrt{H^2_r +R_r/6} } - {2\sqrt{6}\pi \over 3}\right)$. Since we
consider the highly excited strings, i.e. strings with small
momentum $k$, the integral in \rhoappro\ with respect to the
comoving momentum $k$ will contribute a small factor which we have
dropped. This result will not affect the qualitative behavior of the
production rate of strings with respect to the string excitation
modes $n$ and $l$. The upper limit of $n$ is roughly $n_{max} \sim
H^{-2}/4 \gg 1$(See Appendix B), thus $\rho \sim {e^{-A}\over
a(\eta)^3 A^{3/2}}$. We emphasize that indeed $H_r \ll {1\over
\sqrt{\alpha'}}$. Thus in unit where $\alpha' =1$, $H_r \ll 1$ and
$A \gg 1$, the above approximation is qualitatively correct.
Especially, since $A \gg 1$, from \final\ we can see that the energy
density of strings produced is very small and exponentially
suppressed, i.e., highly excited strings are hardly produced in our
case.

\newsec{Discussion}

We have estimated the energy density of strings produced at the end
of inflation and during reheating, our main result is
\eqn\densityfrec{\rho(\eta) \sim {{\alpha'}^{-2}\over a(\eta)^3
A^{3/2}} \left( e^{-A} - {e^{-A\sqrt{n_{max}}} \over n_{max}^{1/ 4}}
\right),}
here $A$ is given by $A=\left({8-12 a_r^2 H_r^2 \alpha'/\pi^2 \over
\sqrt{H^2_r \alpha'+R_r \alpha'/6} } - {2\sqrt{6}\pi \over
3}\right)$. We have reinstated $\alpha'$ which has been set to $1$
in our paper. Although it is difficult to get the explicit form of
$H_r$ due to the complicated equations of \realandimage, one can
make sure that $H_r\sqrt{\alpha'}$ must be much smaller than $1$,
i.e. the Hubble scale in string production process is much lower
than the string scale. In other words, the curvature radius is much
larger than the string length, and a spacetime description gives a
good approximation. In the case of a small $H_r$, due to the large
exponential factor in \densityfrec\ , one can see that the energy
density of strings produced is exponentially suppressed and indeed
highly excited strings are hardly produced.
\kandm. 
Planck scale, otherwise the effective field theory viewpoint we used
will broken down, so that the energy density is also small.

In conclusion, we have shown in this paper that highly excited
strings are hardly produced at the end of inflation, because
$|\beta|^2$ is highly suppressed by a exponential factor within and
the degeneracy of highly excited strings is not sufficiently large
to compensate it, thus the energy density is also suppressed by this
exponential factor.


\noindent{\bf Acknowledgements}

This work was supported by a grant of CNSF. We thank Yi Wang for
discussion.

\appendix{A}{Hubble Parameter During Reheating}

In most of popular inflation scenarios, the temperature is
practically zero during inflation, relativistic matter is produced
during the short reheating period when the inflaton oscillates
coherently and decays to matter. Generally it is not known how the
inflaton is coupled to a generic string state, so the usual
reheating mechanism is not easily applied to the production of
strings.

However, the spacetime metric is also coupled to strings, the
details of the coupling can be seen from the string spectrum
directly. When $H$ remains nearly a constant, there is no string
production. During the reheating period, the Hubble parameter is no
longer a constant, and can be estimated by solving the Friedmann
equation. In solving this equation, we should also take radiation
into account. A more rigorous treatment should also take strings
produced in the process into account, however, we do not know how to
compute string energy density as a function of time (to this end, it
is required to compute the string production rate per unit time).
The Friedmann equation reads
\eqn\friedman{3H^2=\rp+\rg,}
where we set $8\pi G=1$, and $\rp(t)$ and $\rg(t)$ are the energy
densities of the inflaton and radiation respectively, their
equations of motion are
\eqn\eomi{\dot\rp+3H\rp+\g\rp=0,} and
\eqn\eomr{\dot\rg+4H\rg-\g\rp=0,}
where $\g$ is the decay rate of the inflaton and dot denotes the
derivative with respect to the comoving time $t$.

Taking derivative of \friedman\ with respect to $t$ and using \eomi\
\eomr\ to eliminate $\dot\rp$ and $\dot\rg$, and then using
\friedman\ again, we find
\eqn\epsrg{\rg={2\ep-3\over4-2\ep}\rp,}
where $\ep\equiv-\dot H/H^2$. Now the Friedmann equation is
\eqn\epsrgt{3H^2={1\over4-2\ep}\rp .}
Combining \epsrg\ and \epsrgt, we get $1.5<\ep<2$. This result can
be generalized when there are more energy components in the
Universe.

We solve \eomi\ in the limit\foot{Of course one can exactly solve
the equation, but that is not necessary.} of $\g\gg H$, which means
that the inflaton decays very fast, indeed this is the case during
reheating. Thus \eomi\ is simplified to
\eqn\eomil{\dot\rp+\g\rp=0.}  The general solution is
\eqn\eomis{\rp=\rho_0\exp\left(-\g t\right),}
\noindent where $\rho_0$ is an integration constant. Using solution
\eomis, \epsrgt\ can be rewritten as a differential equation of $H$
explicitly
\eqn\equh{6\dot H+12H^2-\rho_0\exp\left(-\g t\right)=0.}
This equation can be cast into the standard form of Bessel equation
by changing variable $t$ into $\tau=\exp(-\g t/2)$,
\eqn\bessl{{d^2(a^2)\over d\tau^2}+{1\over\tau}{d(a^2)\over
d\tau}-{4\rho_0\over3\g^2}(a^2)=0.}
The general solution to this equation is a linear combination of the
modified Bessel functions of the first kind $I_0$ and of the second
kind $K_0$:
\eqn\solubess{a^2=c_1I_0\left(\sqrt{{\rho_0\over3}}{2\tau\over
\g}\right)+c_2K_0\left(\sqrt{{\rho_0\over3}}{2\tau\over \g}\right),}
where $c_1$ and $c_2$ are integration constants.

\appendix{B}{Constraint Equation with Arbitrary $H(t)$}

The equation of motion of bosonic string states derived in \lss\ is
valid in pure de Sitter space, of which the Hubble parameter is a
constant. In order to study string production, we focus on the
reheating phase, where the Hubble parameter is varying with time,
the spectrum formula of \lss\ cannot be used directly for our
purpose. In this appendix we generalize the original result of \lss\
to the case of arbitrary $H(t)\equiv \dot{a(t)}/a(t)$. We refer the
readers to the original paper \lss\ for more details of deriving the
equation of motion when $H$ is a constant.

The key point of the generalization is to re-calculate $c_m$ in
(3.16) of \lss\ in arbitrary $H(t)$ case. The general definition of
$c_m$ is
\eqn\cm{ c_m = e^{i(\phi_m-\psi_m)} \left\{ {\alpha}^{\ast}_m
\dot{\beta}^{\ast}_m
 - {\beta}^{\ast}_m \dot{\alpha}^{\ast}_m + \left[ {{\alpha}^{\ast}}^2 e^{-2i \int^t du \lambda_m(u)} - {{\beta}^{\ast}}^2 e^{+2i \int^t du \lambda_m(u)} \right]{\dot{\lambda}_m\over2\lambda_m}
 \right\},}
where the most general form of $\alpha_m$ and $\beta_m$ are given by
(3.7) and (3.8) of \lss,
\eqn\ab{\eqalign{
    \alpha_m &= \cosh(\gamma_m) e^{i \delta_m + i \phi_m}, \qquad \tilde{\alpha}_m = \cosh(\gamma_m) e^{i \delta_m + i
    \psi_m},\cr
    \beta_m &= \sinh(\gamma_m) e^{i \phi_m}, \qquad \tilde{\beta}_m = \sinh(\gamma_m) e^{i
    \psi_m}.}}
$\lambda_m$ in \cm\ is defined as
\eqn\lambdal{
    \lambda_m \equiv sgn(m)\sqrt{{m^2\over\omega^2} - \eta {\partial}^2_t
    \eta^{-1}},}
where $\eta\equiv {1\over e^{Ht}\sqrt{\omega}}$ as in \lss, and should be replaced by a formula in
which $\exp(Ht)$ is replaced by $a(t)$, and $\gamma_m$ and $\delta_m$ in \ab\ can be solved directly
from (3.9-3.10) of \lss.
\eqn\gammal {
    \gamma_m = \cosh^{-1}{\sqrt{{\omega\over4m\lambda_m}\left[ {\Gamma}^2+\lambda^2_m+{m^2\over\omega^2}
    \right]+{1\over2}}}}
\eqn\deltal {
    \delta_m = \arctan{\left( {-2\lambda_m \Gamma\over2{\Gamma}^2 - \dot{\Gamma}}
    \right)} - 2\int^t du \lambda_m(u).
}
 here we define $\Gamma \equiv \dot{\eta}/\eta$ for short. Insert
\ab\ into \cm\ and use \lambdal-\deltal, after a tedious calculation
we finally get
\eqn\eqcm{
    c_m = e^{-i(\delta_m+\psi_m+\phi_m)}
    \left({\dot{a}\over a}\right){\left[{\ddot{a}\over a}+\partial_t\left({\dot{\omega}\over2\omega}
    \right)-{\left({\dot{a}\over a}\right)}^2 \right] - i {2m\over\omega}\left({\dot{a}\over a}+{\dot{\omega}
    \over2\omega}\right)\over\sqrt{\left[ {\ddot{a}\over a} + \partial_t\left( {\dot{\omega}\over2\omega}\right) -
    {\left({\dot{a}\over a}\right)}^2 \right] + {4m^2\over\omega^2}{\left({\dot{a}\over a}+
    {\dot{\omega}\over2\omega}\right)}}}.
}

What is needed in the calculation of the main context is the modulus of $c_m$
\eqn\result{
    |c_m| = {\dot{a}\over a} \equiv H(t),
}
it agrees with the original result of \lss\ where H is a constant.
\result\ tells us that the equation of motion of
string states in the reheating phase where Hubble parameter is
varying with time $H=H(t)$ is  simply given by the equation of motion of \lss\ when $H$
is replaced by $H(t) =
\dot{a}(t)/a(t)$.

At the end of this appendix we want to recall that the real
condition of each $\lambda_m$ is
\eqn\realc{{m^2\over \w^2}-(H+{\dot\w\over
2\w})^2-\partial_t({\dot\w\over 2\w})>0}
where $\w^2=4N+2E_0+(p^i/a)^2$. In fact if $\lambda_1$ is real, so
are $\lambda_m$ for $m>1$. Then the condition becomes
\eqn\realclI{{1\over \w^2}-\left[1+{\ep\over2}\left({p^i\over a
\w}\right)^2-{3\over 4}\left({p^i\over a \w}\right)^4\right]H^2>0.}
Because
\eqn\reason{\left({p\over a \w}\right)^2={(p^i/a)^2\over
4N+2E_0+(p^i/a)^2 }\ll 1}
and $\ep\in(1.5, 2)$, the condition \realclI\ is approximated by
\eqn\realclI{{1\over \w^2}-\left[1+{\ep\over2}\left({p^i\over a
\w}\right)^2\, \right]H^2>0.}
Thus we get the upper limit of $n$ as
\eqn\upper{n \, < \,
n_{max}={1\over4}\left(H^{-2}-\left(1+{\ep\over2}\right)k^2-2E_0\right).}
where $k^i=p^i/a$ is the physical momentum.

\listrefs

\bye